# Impact of surface roughness on liquid-liquid transition

Ken-ichiro Murata† and Hajime Tanaka*

*Department of Fundamental Engineering, Institute of Industrial Science, University of Tokyo, 4-6-1 Komaba, Meguro-ku, Tokyo 153-8505, Japan*
*† Present address: Institute of Low Temperature Science, Hokkaido University, Kita-19, Nishi-8, Kita-ku, Sapporo, 060-0819, Japan*

* Corresponding author: Email: tanaka@iis.u-tokyo.ac.jp

**Liquid-liquid transition (LLT) in single-component liquids is one of the most mysterious phenomena in condensed matter. So far this problem has attracted attention mainly from the purely scientific viewpoint. Here we report the first experimental study on an impact of surface nano-structuring on LLT by using a surface treatment called rubbing, which is the key technology for the production of liquid crystal displays. We find that such a rubbing treatment has a significant impact on the kinetics of liquid-liquid transition (LLT) of an isotropic molecular liquid, triphenyl phosphite. For a liquid confined between rubbed surfaces, surface-induced barrier-less formation of the liquid II phase is observed even in a metastable state, where there should be a barrier for nucleation of the liquid II phase in bulk. Thus, surface rubbing of substrates not only changes the ordering behavior, but also accelerates the kinetics significantly. This spatio-temporal pattern modulation of LLT can be explained by a wedge filling transition and the resulting drastic reduction of the nucleation barrier. However, this effect completely disappears in the unstable (spinodal) regime, indicating the absence of the activation barrier even for bulk LLT. This confirms the presence of nucleation-growth-type and spinodal-decomposition-type LLT, supporting that LLT is truly a first-order transition with criticality. Our finding also opens up a new way to control the kinetics of LLT of a liquid confined in a solid cell by structuring its surface on a mesoscopic lengthscale, which may contribute to making LLT useful for micro-fluidics and other industrial applications.**

# Introduction

Generally, surface wettability and geometry seriously affects a phase transition near a surface through a coupling of the surface field to the order parameter field. This leads to unique phenomena absent in bulk [1–4]. Such examples can be seen for crystallization, vapor-liquid phase transition [5], and phase separation of a binary mixture [6] in contact with a substrate. Heterogeneous nucleation, i.e., preferential nucleation on a wettable surface, is one of such classical examples, widely seen in our life, ranging from cloud and frost formation in nature to production of high-quality single crystals in industry. These phenomena should not be regarded as a special exception to bulk phase transition since it is almost impossible to avoid the presence of surfaces in any real systems, particularly, in small-size systems relevant to nanotechnology applications. Recently there are growing interests in controlling self-organized pattern evolution by nano-structured functional surfaces using an interplay between phase ordering and wetting [4–9].

Here we focus our attention on how substrate surface geometries affect liquid-liquid transition (LLT) of a single-component liquid, which is the first-order phase transition between two liquid states, liquid I and II. Recently, LLT has attracted considerable attention because of its counter-intuitive nature [10–13]. LLT accompanies distinct change of bulk physical and chemical properties such as density [14], refractive index [14, 15], dielectric constant [16, 17], polarity [18], viscosity [15], fragility [19], and miscibility [20]. Control of these properties while keeping fluidity can be important for practical applications of LLT since the state of a liquid is a key to various transport and chemical reaction processes. LLT offers an intriguing possibility to change the physical and chemical properties of a liquid without modifying its molecular structure. In addition to the change in these bulk properties, we recently found a difference in the wettability between liquid I and II of triphenyl phosphite to a solid substrate [21], suggesting a possibility to control LLT by a surface field. There has been a long-standing debate on the physical nature of the phase transition in triphenyl phosphite [16, 22–34] since its first observation by Kivelson and his coworkers [14, 35]. The difficulty comes from the fact that the transition takes place below the melting point, i.e., in a state intrinsically metastable against crystallization. However, recent experimental studies [18, 36–38] have shown that the transition is LLT and not merely due to nano-crystal formation.

LLT looks counter-intuitive if we assume that a liquid has a completely random structure. However, LLT can be naturally explained by introducing a new additional scalar order parameter [39-42] besides density $\rho$ that is a standard order parameter to describe the liquid state. We argued that this new order parameter should be the fraction of locally favored structures, $S$. Although $\rho$ is a conserved order parameter, $S$ is a non-conserved one since locally favored structures can be created and annihilated independently as spin flipping in magnets. This explain why liquid I completely transforms into liquid II without coexistence, unlike phase separation of a mixture. Then, LLT is regarded as gas-liquid-like cooperative ordering of locally favored structures: liquid I and liquid II are respectively a gas (dilute) and a liquid (dense) state of locally favored structures. Thus, the transformation from liquid I to II is the process of the increase in the fraction of locally favored structures [37]. In this scenario, a gas-liquid transition is dominated by $\rho$ whereas a liquid-liquid transition is dominated by $S$.

Here we report the first experimental study on an impact of surface nano-structuring on LLT. By comparing the ordering process of LLT of triphenyl phosphite in contact with rubbed substrates to that with unrubbed smooth ones for both nucleation-growth (NG)-type and spinodal-decomposition (SD)-type LLT, we find that surface structuring by rubbing, which is widely used in liquid-crystal industries, strongly accelerates the kinetics of NG-type LLT. We show that rubbing significantly affects the pattern formation and accelerates the kinetics for NG- type LLT, but has no effects on SD-type LLT. This sheds new light on the very nature of LLT. Furthermore, this controllability of the kinetics of LLT by wedge wetting effects will open a new avenue of LLT research towards micro-fluidics and other industrial applications.

## Results and Discussion
### Rubbing effects on nucleation-growth-type LLT

First we show, in Fig. 1A, a typical pattern evolution process of LLT in a cell made of unrubbed planar surfaces, which was observed with phase contrast microscopy at the annealing temperature $T_a$ = 218 K (see also movie S1). This temperature is located in the metastable region of LLT, or below the binodal temperature $T_{BN}$ but above the spinodal temperature $T_{SD}$ = 215.5 K. In an unrubbed cell, spherical droplets of liquid II are nucleated and increase their size linearly with time in the matrix of liquid I, and finally fill up an entire space. The schematic explanation of NG-type pattern evolution in a system confined between two solid substrates is also shown in Fig. 1A. On the other hand, in a rubbed cell, whose surfaces have a plenty of submicron-scale grooves formed along the rubbing direction (see the white arrow), barrier-less formation of liquid II domains was observed (see Fig. 1B and movie S2). In this case, LLT is initiated by the barrier-less formation of liquid II domains on the rubbed substrates, which is followed by the growth towards the middle of the cell, resulting in the gradual increase of the intensity of the image with time, unlike the above case of typical NG-type pattern evolution, where dark droplets of liquid II are directly formed in the middle of the matrix of liquid I (see the schematic pictures in Fig. 1). As shown below, this leads to a difference in the temporal change in the intensity distribution (see Fig. 2). This behavior is fundamentally different from the ordinary NG-type transformation shown in Fig. 1A, despite that both processes take place at the thermodynamically identical condition. However, even in Fig. 1B, we can see the nucleation of dark droplets (see, e.g., the circled one) in the intermediate stage of the process in addition to the pattern formed near the surfaces. We found that these dark droplets of liquid II do not appear for a cell whose thickness is less than 5 $\mu$m. Together with microscopy observations with different focal depths, we conclude that nucleation of dark liquid II droplets overcoming a barrier takes place in the middle (bulk) part of the cell after the surface-assisted barrier-less formation of the liquid II domains takes place (see the schematic picture of Fig. 1A). We note that in the late stage these droplets are merged with the liquid II wetting layers formed on the walls which propagate from the surfaces towards the middle of the cell, and thus eventually merges with the droplets. We confirm that the surface effects are significant at least for a cell thickness less than a few tens of $\mu$m.

We also measured the temporal change in the intensity distribution $P(I)$ of a microscopy image, which is proportional to the density distribution $P(\rho)$. Note that phase contrast microscopy detects the difference in the refractive index, which is proportional to the density difference: a darker contrast means a higher density in our measurements [43]. Since the locally favored structure has a higher density than the normal-liquid structure, $P(\rho)$ reflects $P(S)$. As shown in Fig. 2A, the temporal change of $P(I)$ for the unrubbed cell (see Fig. 1A) is identical to that of our previous study [44]: right large and left small peaks, respectively corresponding to liquid I (bright matrix region) and liquid II (dark droplets), coexist in the intermediate stage of LLT, which is characteristic of NG-type transformation. It is well known that, in nucleation-growth-type ordering, nuclei with the final order parameter value (forming the peak of low intensity in this case) appear in the majority matrix (forming the peak of high intensity) and the peak of nuclei grows without a big change in the peak position (or, the order parameter value). Here we note that the left small peak is very broad due to a focal depth problem (see Supplementary Materials for the detail). In contrast, $P(I)$ for the rubbed cell (see Fig. 1B) does not have a bimodal-like shape but rather broadens while changing its peak position continuously from the right peak of liquid I to the left peak of liquid II without creating another peak (see Fig. 2B). In the rubbed cell, the growth of the liquid II phase is not limited by the nucleation process since there is no barrier for its formation, but mainly by the growth process towards the middle of the cell. Thus, the continuous increase of the wetting layer thickness is the origin of the continuous shift of the peak intensity (see the schematic picture in Fig. 1B). This temporal change of $P(I)$ is reminiscent of the behavior of SD-type LLT [44]. This similarity comes from the barrier-less formation of the liquid II phase, but at the same time there is a crucial difference in the origin of the continuous change in the peak intensity: For ordinary SD-type LLT the order parameter itself changes continuously with time, whereas in the above case the continuous nature comes from the fact that the intensity is proportional to the thickness of the wetting layer, which increases continuously with time. Note that the intensity of the image is the integration along the thickness direction. Furthermore, we can see clearly by comparing Fig. 1A and B that not only the spatial pattern but also the kinetics of LLT is significantly accelerated by surface rubbing, due to the barrier-less ordering.

To see this more quantitatively, we performed dielectric spectroscopy measurements and evaluated the time evolution of the normalized dielectric strength of liquid II, $\Delta\varepsilon_{II}(t)/\Delta\varepsilon_{II}$, which is a measure of the degree of the transformation from liquid I to liquid II. Figure 2C shows the time evolution of $\Delta\varepsilon_{II}(t)/\Delta\varepsilon_{II}$ at $T_a = 219$ K and 220 K for a sample sandwiched between rubbed substrates together with that between unrubbed smooth ones. We found that, for both temperatures, the growth of $\Delta\varepsilon_{II}(t)/\Delta\varepsilon_{II}$ in the rubbed cell is accelerated two- or three-fold compared to that in the unrubbed one. As will be discussed below, LLT in a rubbed cell proceeds as if there is no activation barrier even in a metastable state above $T_{SD}$. Since the thermodynamic condition is identical and accordingly the structural relaxation time is the same between the two cases, the growth kinetics of the order parameter should also be the same. Thus, we conclude that the dynamical acceleration of LLT is a consequence of the absence of the activation barrier for nucleation of liquid II droplets on rubbed surfaces: surface rubbing induces spontaneous barrier-less growth of the order parameter even in a metastable state of the bulk liquid.

**Origin of rubbing-induced acceleration of nucleation-growth-type LLT**

Next we consider the physical origin of the spatio-temporal modulation of LLT induced by rubbing. Figure 3A shows a confocal microscopy image of a rubbed substrate. We confirmed a large number of microgrooves aligned along the rubbing direction, with typical intervals of a few $\mu$m. Figure 3B shows a transient spatial pattern of LLT at $T_a = 220$ K. There we can see complete filling of microgrooves by liquid II as well as partial filling by liquid II droplets on aligned microgrooves, which is the origin of the anisotropy in the two-dimensional (2D) power spectrum pattern (identical to the light scattering pattern [45]) shown in Fig. 3C. There we can see a bright line at $q_X = 0$ in addition to the ordinary isotropic spinodal pattern, indicating the presence of nearly one-dimensional liquid II domains along the rubbed direction (or, on microgrooves).

The ordering behavior of liquid II on a rubbed substrate is reminiscent of the so-called wedge filling of a liquid [46], provided that a microgroove can be regarded as a wedge. The wetting morphology for a wedge is known to be classified into three types in terms of the relation between the contact angle to the surface, $\theta$, and the wedge angle, $\alpha$ [47]: For $\theta < 90°-\alpha/2$, a liquid completely fills the wedge with a negative Laplace pressure. For $90-\alpha/2 < \theta < 90°+\alpha/2$, a liquid partially fills the wedge in a droplet-like form with a positive Laplace pressure. For $\theta > 90°+\alpha/2$, no wedge filling takes place. We mention that the last condition is satisfied for a water/glycerol mixture, which also exhibits LLT [17]. For this system, we did not see any indication of surface-assisted modulation of LLT, but only observed a usual NG-type process similar to that in Fig. 1A. This difference in the behavior between triphenyl phosphite and a water/glycerol mixture is a consequence of the rather low wettability of liquid II of the latter to the glass substrate compared to the former.

Our two-order-parameter model of liquids [39-41] suggests that there should be similarity in the pattern evolution between LLT influenced by a rubbed surface and a gas-liquid phase transition under the influence of wedge filling [5, 9, 47] (see Introduction). Indeed, we recently demonstrated that, in the process of LLT of triphenyl phosphite in contact with a solid wall, a transition from partial to complete wetting of liquid II takes place on an unrubbed planar substrate when approaching $T_{SD}$ [21], which is reminiscent of "critical point wetting" near a gas-liquid critical point [1, 3]. As schematically shown in Fig. 4, the decrease of $\theta$ near $T_{SD}$ and $T_W$ should also lead to a transition from partial to complete filling (wedge filling transition) at a filling temperature $T_F$, which satisfies $\theta(T_F) = 90° - \alpha/2$. We confirmed that this transition takes place around 220 K: $T_F \sim 220$ K (see also movie S3). Note that distributions (or, randomness) of both the wedge angle and the length of microgrooves produced by rubbing may make the transition rather broad.

Nucleation of droplets basically proceeds with the formation of a new interface between liquid I and liquid II, whose profile is determined to locally minimize the free energy [48]. For a system satisfying the filling condition ($T < T_F$, or $\theta < 90° - \alpha/2$), liquid II droplets nucleate and grow while filling a wedge in liquid I. In the critical regime ($T \sim T_F$, or $\theta \sim 90° - \alpha/2$), on the other hand, nucleation of liquid II on a wedge is expected to accompany contact-line fluctuations induced by critical wedge filling. Furthermore, the kinetics of LLT on rubbed substrates is distinct from the usual NG-type transformation, which is the activation process overcoming a free energy barrier associated with the formation of a new interface. For $\alpha/2 \leq 90° - \theta$, such a barrier for the formation of a new phase vanishes selectively on wedges, leading to a situation similar to the formation of a new phase on a complete wettable surface. Then, the formation of liquid II on a rubbed substrate should proceed spontaneously without an activation barrier, despite the fact that the system is still in a thermodynamically metastable state and not in an unstable state (above $T_{SD}$). We argue this is the origin for the kinetic acceleration of LLT in a rubbed cell (see Fig. 2C). We tried to obtain the detailed geometrical characteristics of the surface pattern produced by rubbing with atomic force microscopy (AFM). However, it was difficult to characterize the angle $\alpha$ because of the random nature of rubbed surfaces and artifacts coming from the shape of the probe of AFM (see Supplementary Materials). To make a quantitative check, we need to use well-controlled surface structures as done in Ref. [49], which remains for future research.

**Rubbing effects on spinodal-decomposition-type LLT.**

Finally, we consider wedge wetting effects on SD-type LLT. Figure 5A shows a spatial pattern evolution process of SD-type transformation on a rubbed surface at 214 K. We also show a direct comparison of the evolution of $\Delta\varepsilon_{II}$ between a rubbed and an unrubbed cell in Fig. 5B. Unlike the NG case, there are no effects of rubbing on pattern evolution. The patterns are isotropic (see also the isotropic 2D power spectrum images shown in the insets of Fig. 5A) and there is no kinetic acceleration. This behavior is fully consistent with our previous observation that surface wetting effects have no impact on SD-type LLT even for a complete-wetting surface [21]. The absence of wedge and surface wetting effects is a direct consequence of (i) short-range nature of the surface field [21] and (ii) the non-conserved nature of the order parameter [40]. The former leads to an extremely slow growth of wetting (or filling) layer ($\sim \ln t$) for an unstable regime [50] compared to bulk coarsening. Thus, fast-growing fluctuations in bulk overwhelm slowly-growing wetting layers. This is because the latter allows only a local change of the order parameter near the surface and, thus, cannot be delocalized due to the irrelevance of diffusion and/or flow unlike in a system of a conserved order parameter [51]. This is a consequence of the non-conserved nature of the order parameter, which can change locally without accompanying diffusion. For a system of a conserved order parameter (e.g., phase separation of a binary mixture and a vapor-liquid transition), on the other hand, wedge wetting effects are expected to have a serious influence on the kinetics of SD-type transformation via surface-induced composition waves and interfacial tension driven flow, reflecting its non-local nature [51].

In the above we have shown that the effects of rubbing on LLT are significantly different between below and above $T_{SD}$. Figure 5C shows the ratio of the incubation time of LLT in an unrubbed cell, $\tau_0$, to that in a rubbed cell, $\tau_R$, below $T_F$. Here we independently determine the spinodal point $T_{SD}$ by optical microscopy measurements [15] (see the vertical dashed line). Although the accuracy of the determination of the ratio, $\tau_0/\tau_R$, is not so high (see the error bars in Fig. 5C), we can see that, with decreasing the temperature, $\tau_0/\tau_R$ drastically decreases towards $T_{SD}$ in the metastable (NG) region, and then becomes a constant ($\tau_0/\tau_R = 1$) in the unstable (SD) region, which means no kinetic acceleration below $T_{SD}$. This strongly supports the above-described scenario. The relation $\tau_0/\tau_R=1$ suggests that the bulk growth is dominant even in a rubbed cell below $T_{SD}$ (see above). Note that, as the system approaches $T_{SD}$, the nucleation barrier for liquid II droplets decreases and thus rubbing effects gradually become less important. Accordingly, the difference in the kinetics between the rubbed and unrubbed cases gradually become smaller towards $T_{SD}$ and finally disappears completely at $T_{SD}$.

## Conclusion

In conclusion, we have demonstrated that the spatio-temporal pattern evolution of NG-type liquid-liquid transition in triphenyl phosphite is seriously influenced by surface rubbing, which induces a wedge filling transition. Even in a metastable region, rubbing induces spontaneous barrier-less transformation from liquid I to II on the substrates and thus accelerates the kinetics of liquid-liquid transition significantly. We also show that the absence of wedge wetting effects on the kinetics of SD-type liquid-liquid transition is unique to the ordering governed by a non-conserved order parameter. We note that this is markedly different from a system of a conserved order parameter (e.g., a binary liquid mixture), where surface wetting effects seriously affect the pattern evolution of spinodal decomposition and accelerates the kinetics via diffusion or hydrodynamic flow [51].

Furthermore, the disappearance of wedge wetting effects below the spinodal temprature $T_{SD}$ strongly suggests that there is no activation barrier for the transformation of liquid I to II below $T_{SD}$, i.e., the transition mode really changes from NG-type (metastable) to SD-type (unstable) liquid-liquid transition at $T_{SD}$. The presence of SD-type ordering further supports that the transition is primarily liquid-liquid transition and 'not' nano-crystal formation (see also [18, 36-38, 41, 42] and the references therein). As mentioned above, the behavior also indicates that the order parameter governing the liquid-liquid transition is of non-conserved nature. These findings shed new light on the physical nature of liquid-liquid transition, including the characterization of the nature of the order parameter governing it.

Finally, from an applications viewpoint, our finding may open up a novel way to control the kinetics of liquid-liquid transition in a metastable state, which leads to a change in liquid properties such as density, refractive index, chemical properties, and transport properties, by using surface structuring of a container. Such surface effects should be particularly important for nano- and micro-fluidics applications. Furthermore, the development of surface fabrication techniques capable of providing designed nano-structures would offer more accurate spatio-temporal control of liquid-liquid transition.

**Acknowledgments**

The authors are grateful to John Russo for a critical reading of the manuscript and Ken Nagashima for his technical supports of AFM measurements. This work was partially supported by Grands-in-aid for Scientific Research (S) and Specially Promoted Research from the Japan Society for the Promotion of Science (JSPS).

## Materials and Methods

**Experimental details**

The sample used in this study is triphenyl phosphite, which was purchased from Acros Organics. We used it after extracting only a crystallizable part to remove impurities. We carefully avoided moisture to prevent chemical decomposition of triphenyl phosphite by using dried nitrogen gas, since the molecules are known to decompose under the presence of water because of ester hydrolysis. The temperature was controlled within ±0.1 K by a computer-controlled hot stage (Linkam LK-600PH) with a cooling unit (Linkam L-600A). Sample cells used in this measurement are purchased from E.H.C. Co Ltd, which are composed of two parallel glass substrates coated with three or ten times rubbed polyimide, which we call rubbed cells, and with unrubbed polyimide, which we call usual cells. For the rubbed cells, both top and bottom surfaces are rubbed along the same direction. We employed a Leica SP5 confocal microscope to measure a surface roughness pattern of the rubbed substrate. An Indium Titan Oxide (ITO) transparent electrode is also deposited on the glass substrate, which allows us to perform simultaneously dielectric spectroscopy measurements and microscopy observation. Here the transparent electrode has no effect on the wetting dynamics of the LLT since polyimide is further coated on the electrodes. The cell thickness is fixed to 5 or 10 $\mu$m with an error of ±1 $\mu$m. We observed a transformation process from liquid I to liquid II with phase contrast microscopy (Olympus, BH2-UMA). Broadband dielectric measurements were performed in the frequency range from 10 mHz to 1 MHz with an Impedance/Gain-Phase Analyzer (Solatron SI1260).

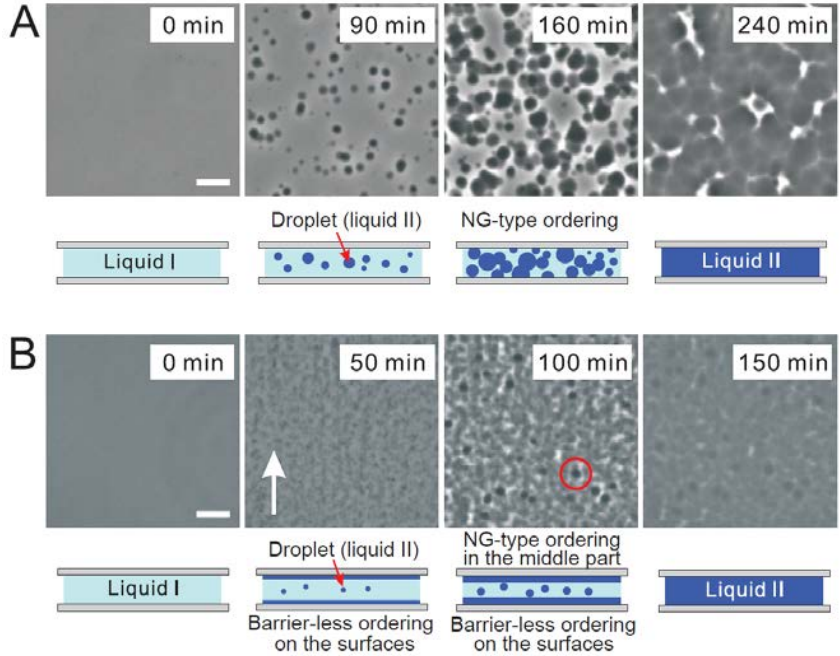

**Fig. 1. Comparison of pattern evolution between LLT on smooth unrubbed and rubbed surfaces.** Pattern evolution observed at $T_a$=218 K in an unrubbed smooth (**A**) and a rubbed cell (**B**) with phase contrast microscopy. The pattern formation process can be seen also in movie S1 and S2 for case A and B, respectively. The cell thickness is 10 $\mu$m for both cases and the scale bars correspond to 20 $\mu$m. Schematic figures illustrate cross-sectional views of the ordering processes.

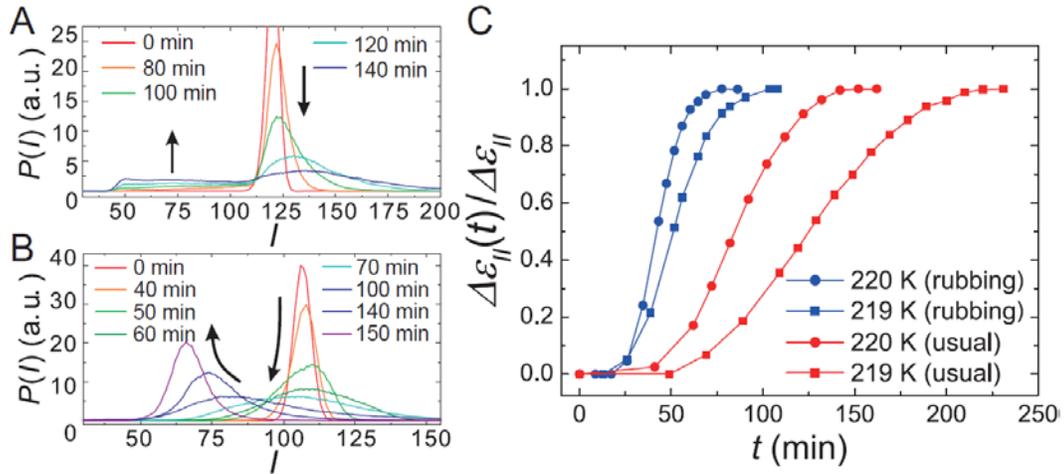

**Fig. 2. Dynamic acceleration of LLT on rubbed surfaces relative to that on unrubbed ones.** Temporal change of $P(I)$ during LLT at $T_a$=218 K for a smooth (**A**) and a rubbed (**B**) cell. (**C**) Time evolution of the normalized dielectric relaxation strength of liquid II, $\Delta\varepsilon_{II}(t)/\varepsilon_{II}$, at $T_a$=219 K (squares) and 220 K (circles) for rubbed (blue) and unrubbed (red) cells. It is found that, for both temperatures, the evolution of $\Delta\varepsilon_{II}$ becomes two or three times faster in rubbed cells than in usual cells.

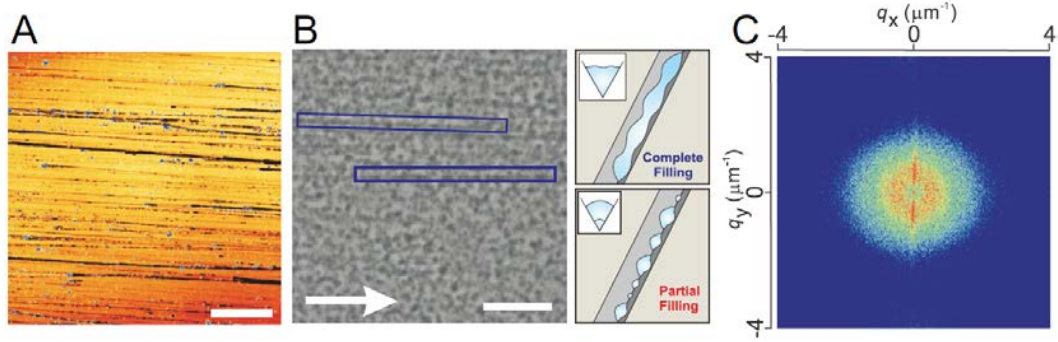

**Fig. 3. Anisotropic wetting morphologies induced by rubbing.** **(A)** A confocal micrograph of the 10-times rubbed substrate. The color in the image corresponds to the height level: The planer area appears yellow. The black color indicates height lower than the planer surface, corresponding to scratches (grooves) formed by rubbing, whereas the blue color indicates height higher than the planer surface, corresponding to ridges or impurities on the substrate. Note that the depth of microgrooves is less than the resolution of optical and phase contrast microscopy. Because of the random nature of grooves, their precise characterization by atomic force microscopy was difficult. **(B)** A transient pattern observed with phase contrast microscopy during LLT on the 10-times rubbed substrate at $T_a$=220 K. The white arrow indicates the rubbing direction (i.e., the $x$-direction). The complete filling of a wedge by liquid II appears as a filament-like shape (see the regions enclosed by the blue boxes), whose situation is depicted in the upper-right schematic figure. In addition to this, the partial filling by liquid II, whose situation is depicted in the lower-right schematic figure, is also observed. Such coexistence of various wetting patterns may be a consequence of distributions of the wedge angle, the depth, the length of grooves formed by rubbing. **(C)** A 2D power spectrum pattern of the image in B (see [45] for the calculation method). The scale bars in **(A)** and **(B)** correspond to $50\mu$m.

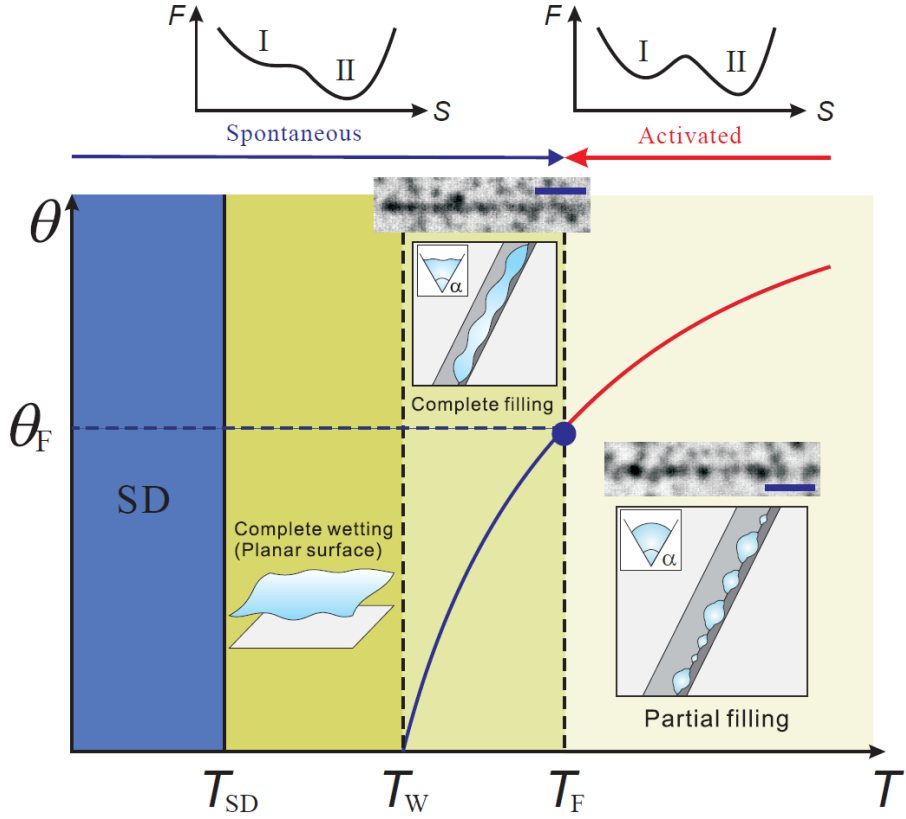

**Fig. 4. Relationship between the wedge filing transition, wetting transition, and spinodal point of LLT.** On the top, we show schematic figures of the free energy $F$ for two situations, where liquid I is unstable (left) and metastable (right) against liquid II. In our system, the contact angle of liquid II decreases towards zero when approaching $T_{SD}$, since the interfacial tension vanishes towards $T_{SD}$=215.5 K [21]. Thus, a transition from partial to complete wetting (wetting transition) takes place at the wetting temperature $T_W$ above $T_{SD}$. For a wedge whose angle is $\alpha$, the decrease of $\theta$ near $T_{SD}$ and $T_W$ also induces a transition from partial to complete filling (wedge filling transition) at the filling temperature $T_F$, where the condition $\theta(T_F)=90°-\alpha$ is satisfied. The two phase contrast microscopy images indicate the formation of the liquid II phase on a well-developed wedge at 218 K and 220 K, respectively. The black scale bars there correspond to 10 $\mu$m. Partially filled liquid II begins to appear in the form of droplets above 220 K. The filling temperature $T_F$ of our system is thus considered to be located around 220 K. Note that, in the rubbed cell we used, the surface of the substrate was coated by polyimide, whose contact angle was estimated as 81° at 220 K from our previous study [21]. The wedge angle of the microgrooves is thus estimated as 9°. Because of the high contact angle, $T_W$ is expected to be located near the spinodal temperature $T_{SD}$, i.e., $T_W \sim T_{SD}$, which makes it very difficult to experimentally determine the wetting transition $T_W$ on a planar polyimide surface.

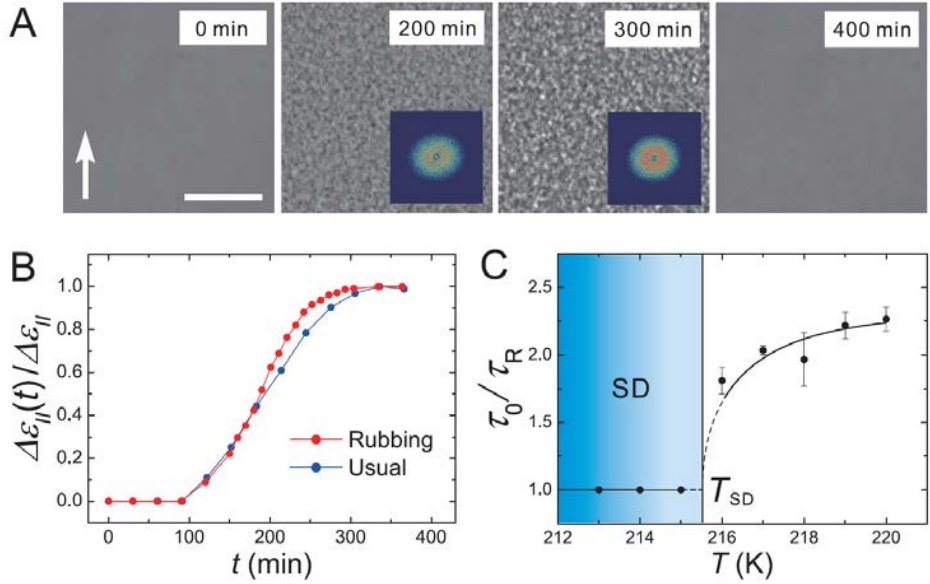

**Fig. 5. Pattern evolution during SD-type LLT on a rubbed substrate.** (**A**) Pattern evolution observed with phase contrast microscopy at $T_a = 214$ K. We observed the initial growth of the amplitude of the order-parameter ($S$) fluctuations through its coupling to the density (at 200 min) and its coarsening in the later stage (at 300 and 400 min) during LLT. This pattern evolution is identical to that of typical SD-type LLT in triphenyl phosphite [15, 44]. The white arrow indicates the rubbing direction. The scale bar corresponds to 50 $\mu$m. Insets (at 200 min and 300 min) are 2D power spectrum images obtained by digital image analysis [40], which is isotropic unlike the case of NG-type LLT (see Fig. 3C). (**B**) A direct comparison of $\Delta\varepsilon_{II}(t)/\Delta\varepsilon_{II}$ at $T_a$=214 K between the rubbed and smooth unrubbed surface case. (**C**) Temperature dependence of the ratio of the incubation time of LLT between the rubbed surface case, $\tau_R$, and the smooth unrubbed surface case, $\tau_0$. We define $\tau_R$ and $\tau_0$ as the time when a SD-like pattern appears on the surface and when the area fraction of droplets of liquid II in liquid I matrix reaches 0.02, respectively. LLT on the rubbed surface proceeds spontaneously without an activation energy below $T_F$, whereas LLT in bulk does so only below $T_{SD}$. Above $T_{SD}$, LLT in the unrubbed surface case (or, in bulk) proceeds by overcoming an activation barrier required for nucleation of liquid I droplets.

# Supplementary Materials for
## "Impact of surface roughness on liquid-liquid transition"

Ken-ichiro Murata and Hajime Tanaka

**THE SHAPE OF THE INTENSITY DISTRIBUTION FUNCTION** $P(I)$

In the main text, we state that we should have a double peak shape for nucleation-growth (NG)- type LLT. Actually, however, we do not see a clear double peak feature in Fig. 2A. This is because observation of droplets smaller than the focal depth by phase-contrast microscopy results in a weaker contrast of droplets due to defocusing effects. As shown in fig. S1, we can clearly see double peaks in $P(I)$ at higher temperatures $T$ = 219 and 220 K, but not at $T$ = 218 K, where the droplet size is smallest among the three temperatures. Note that with a decrease in the annealing temperature, the number of nuclei steeply increases and thus droplets are merged before becoming big enough compared to the sample thickness. So, although $P(I)$ does not show a clear low intensity peak, we conclude together with direct optical microscopy observation (see Fig. 1A), that the pattern evolution observed at 218 K of TPP confined in an unrubbed cell belongs to NG- type LLT.

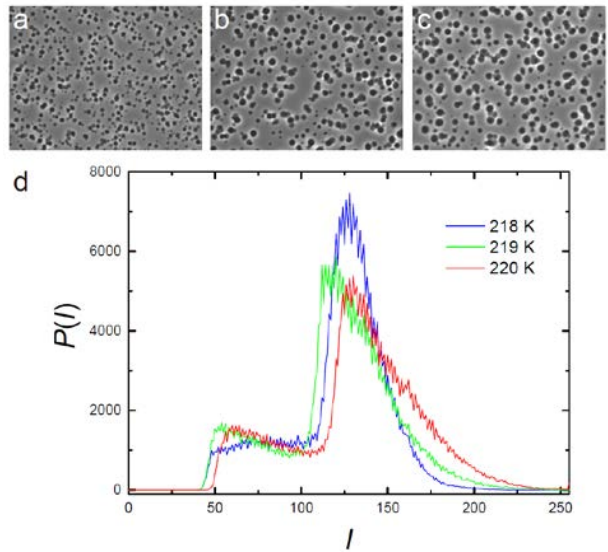

Figure S1. **Relation between a phase contrast image and the intensity distribution function.** Phase-contrast microscopy images observed at 115 min for 218 K (a), at 100 min for 219 K (b), and at 75 min for 220 K (c). The image size is 288 ×216 $\mu$ m. (d) Intensity distribution functions $P(I)$ calculated from these three images.

**THE AFM OBSERVATION OF THE RUBBED SUBSTRATE**

In addition to laser confocal microscopy (Fig. 3A), we also evaluate the surface topography of the rubbed substrate with atomic force microscopy (AFM) (contact mode AFM; Bruker Dimen- sion 3100). Such an example is shown in fig. S2. In the AFM observation, we confirmed the microgrooves aligned along the rubbed direction, as shown in Fig. 3A. However, the quantitative evaluation of the microgrooves suffers from artifacts coming from the shape of the probe of AFM. The half-cone angle of the tip of the cantilever is 25° and the tip height is 17 $\mu$m. We note that the former value is significantly larger than our estimated half-wedge angle of the grooves, $\alpha/2 = 9°$, which hampers the probe to access the tips of narrow and sharp grooves and makes the precise evaluation of them impossible.

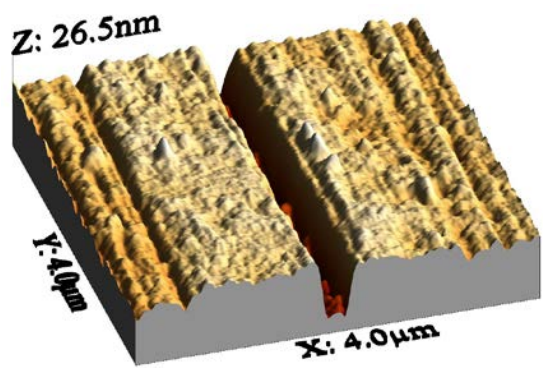

Figure S2. **AFM image of the rubbed surface.** An example of surface topography observed by AFM. We can see microgrooves, but their surface topography suffers from artifacts coming from the shape of the AFM probe.